# Direct Probing of 1/$f$ Noise Origin with Graphene Multilayers: Surface *vs.* Volume

Guanxiong Liu[1], Sergey Rumyantsev[2,3], Michael S. Shur[2] and Alexander A. Balandin[1,4,*]

[1]Nano-Device Laboratory, Department of Electrical Engineering, Bourns College of Engineering, University of California – Riverside, Riverside, California 92521 USA

[2]Center for Integrated Electronics and Department of Electrical, Computer and Systems Engineering, Rensselaer Polytechnic Institute, Troy, New York 12180 USA

[3]Ioffe Physical-Technical Institute, The Russian Academy of Sciences, St. Petersburg, 194021 Russia

[4]Materials Science and Engineering Program, University of California – Riverside, Riverside, California 92521 USA

*Correspondence to:  balandin@ee.ucr.edu

**Abstract**: Low-frequency noise with the spectral density $S(f) \sim 1/f^{\gamma}$ ($f$ is the frequency and $\gamma \approx 1$) is a ubiquitous phenomenon, which hampers operation of many devices and circuits. A long-standing question of particular importance for electronics is whether $1/f$ noise is generated on the surface of electrical conductors or inside their volumes. Using high-quality graphene multilayers we were able to directly address this fundamental problem of the noise origin. Unlike the thickness of metal or semiconductor films, the thickness of graphene multilayers can be continuously and uniformly varied all the way down to a single atomic layer of graphene – the actual *surface*. We found that $1/f$ noise becomes dominated by the volume noise when the thickness exceeds ~7 atomic layers (~2.5 nm). The $1/f$ noise is the surface phenomenon below this thickness. The obtained results are important for continuous downscaling of conventional electronics and for the proposed graphene applications in sensors and communications.

**Article Text:** Low-frequency noise with the spectral density $S(f) \sim 1/f^{\gamma}$ ($f$ is the frequency and $\gamma \approx 1$) is a ubiquitous phenomenon, first discovered in vacuum tubes (*1*) and later observed in a wide variety of electronic materials (*2-5*). The importance of this noise for electronic and communication devices motivated numerous studies of its physical mechanisms and methods for its control (*6*). However, after almost a century of investigations, the origin of $1/f$ noise in most of material systems still remains a mystery. A question of particular importance for electronics is whether $1/f$ noise is generated on the surface of electrical conductors or inside their volumes. Here we show that by using mechanically exfoliated high-quality graphene multilayers one can address *directly* the fundamental problem of the surface *vs.* volume origin of $1/f$ noise. Unlike the thickness of metal or semiconductor films, the thickness of graphene multilayers can be continuously and uniformly varied all the way down to a single atomic layer of graphene – the actual *surface*. Using a set of samples with the number of atomic planes, *n*, varying from $n=1$ to $n=15$, we found that in moderately-doped samples 1/*f* noise becomes dominated by the *volume noise* when the thickness exceeds $n≈7$. The latter is an unexpected discovery considering that seven atomic layers constitute a film of only ~2.5-nm thickness. Our results reveal a scaling law for the 1/*f* noise, which is important for nanoscale devices and for proposed graphene applications in sensors, analog circuits, and communications.

**The problem of 1/*f* noise origin:** It is hard to find another scientifically and practically important problem that has ignited so many debates but remained unsolved for almost a century as the problem of volume *vs*. surface origin and mechanism of 1/*f* noise in electrical conductors (*7-16*). The intensity of discussions can be inferred even from the titles of seminal publications on the subject: "1/*f* noise is no surface effect" (1969) (*9*) followed by "1/*f* noise: still a surface effect" (1972) (*10*). The direct test of whether measured 1/*f* noise is dominated by contributions coming from the sample surface or its volume has not been possible because of inability to fabricate continuous metal or semiconductor films with the uniform thickness below ~8 nm (*14, 16*). The state-of-the-art in noise field is characterized by existence of a large number of *ad hoc* models each tailored to a specific material system or a device. For example, one of few conventionally accepted theories – McWhorter model (*7*) – deals with 1/f noise in field-effect transistors. The fundamental understanding of the origin of 1/*f* noise in homogeneous electrical conductors is still lacking.

The recent advent of the mechanically exfoliated graphene, which allowed for investigation of its electronic (*17-18*) and thermal (*19*) properties, opens up a possibility of revisiting the old problem of the 1/*f* noise origin. Owing to their layered crystal structure and presence of the van-der-Waals gaps, the multilayer graphene films can be made nearly defect-free and *continuous* down to the thickness of a single atomic layer – an ultimate surface. The noise response in graphene multilayers is not affected as much by the grains and roughness like in metals. The thickness of the graphene multilayers can be made uniform along their entire area and determined with better accuracy than that of the metal films or conducting channels in the semiconductor transistors used previously for the noise studies. Observing the evolution of the noise-power spectral density $S_I(f)$ as the number of atomic planes gradually decreases from a relatively thick graphite film to a single layer graphene (SLG) one can answer the fundamental question of whether the noise is coming from the volume or surface or both and, if 1/*f* noise is coming from both the surface and volume – at what thickness of the conductor the volume contribution becomes dominant. The knowledge of the thickness at which the 1/*f* noise is the *surface phenomenon* has important implications for downscaling electronics beyond a few nanometers.

**Sample Preparation and Measurements:** The graphene multilayers under study were produced by the mechanical exfoliation on $Si/SiO_2$ substrate (*17*). The back-gated devices were fabricated by the electron-beam lithography with Ti/Au (6-nm/60-nm) electrodes (*20*). Micro-Raman spectroscopy is an excellent tool for determining the number of atomic plans for *n*<5 (*21-22*). For larger *n* – essential for this study – the method becomes ambiguous. For this reason, we combined Raman spectroscopy with atomic force microscopy (AFM). The details of *n* value extraction are given in *Supplementary Information*. Figure 1a-d shows scanning electron microscopy (SEM) and AFM images of a typical graphene multilayer device under test, evolution of the Raman spectra as $n$ changes from 1 to 15 and the sheet resistance, $R_{ST}$, vs. back-gate bias for different multilayers. The absence of the disorder $D$ peak in Raman spectra confirms a high quality of our samples (*22*). Since the gating is needed for determining the charge neutrality point (CNP) in different samples, we limited the thickness of the films to $n$=15 where the gating becomes weak.

[Figure 1]

The noise measurements were performed using conventional instrumentation (*Supplementary Information*). Figure 2a shows typical noise spectra for graphene multilayers with different $n$. In all cases, the noise spectral density was close to ~$1/f^{\gamma}$ with $\gamma\approx1$. The absence of Lorentzian bulges indicates that no trap with a specific time constant dominated the spectra (*6, 20*). The $S_I$ proportionality to $I^2$ (Figure 2b) implies that the current does not drive the fluctuations but merely makes them visible as in other homogeneous conductors (*16*). It is informative to correlate the normalized noise spectral density $S_I(f)/I^2$ with the sample resistance. For proper comparison, $S_I(f)/I^2$ should also be normalized to the graphene device area $A$ (*23-24*). In our samples, $A$ varied from 1.5 to 70 μm$^2$. Figure 2c shows $S_I(f)/I^2 \times A$ vs. $R_{ST}$ in graphene multilayers near CNP at $f$=10 Hz. The experimental data can be fitted with two segments of a different slope. At the low resistance range – corresponding to thicker multilayers – the $1/f$ noise is proportional to $R_{ST}$ while at the higher resistance range – thinner multilayers – the dependence is close to $R_{ST}^2$. For homogeneous metals and semiconductors, $S_I(f)$ proportionality to $R_{ST}$ or $R_{ST}^2$ was interpreted as indication of the volume or surface noise origin, respectively (*16*).

It was previously established that 1/$f$ noise in the single layer graphene (SLG) devices has larger contribution from the graphene-metal contacts that graphene multilayer devices (*25*). The deposition of metal contacts to graphene results in graphene doping via the charge transfer to reach the equilibrium conditions, and, correspondingly, leads to the local shift of the Fermi level in graphene (*26*). The electron density of states in graphene near CNP is low owing to the Dirac-cone linear dispersion. Thus, even a small amount of the charge transfer from or to the metal affects the local Fermi energy of graphene stronger than in graphene multilayers. In Figure 2c we indicated a data point for a sample which had a single atomic plane thickness along the conducting channel region but gradually increased thickness (to $n=2$ or 3) under the metal contacts. This sample can be viewed as SLG with reduced graphene-metal contact contribution to the noise. One can see that the data point corresponding to this device fits the $R_{ST}^2$ dependence better. Figure 2d presents the dependence of $S_I(f)/I^2 \times A$ on the gate bias, $V_G$, for multilayers with different *n*. As expected the noise amplitude is the highest in SLG (*27-28*). The noise decrease with *n* is consistent with the sheet resistance change presented in Figure 2c. The wide range of the back-gate bias allowed us to probe the noise near CNP and in the high-bias regime ($|V_G|>40$ V) characterized by the large carrier densities.

[Figure 2]

**Noise scaling and its origin:** The electrical conductance in uniform metal and semiconductor channels follows the Ohm's law: $R=\rho \times (L/W \times H) = R_{ST} \times (L/W)$, where $\rho$ is the resistivity, $L$ is the length, $W$ is the width and $H$ is thickness of the conductor. However, $R_{ST} \sim 1/H$ scaling is not necessarily valid for graphene multilayers as $H$ approaches a single atomic plane – surface.

Figure 3a shows measured $R_{ST}$ as a function of $H$. Our data agree well with independent studies of $R_{ST}(H)$ (29). The linear fit in the logarithmic plot works well for $n$ decreasing from 15 to 2. The $R_{ST}$ value for SLG falls off this dependence suggesting that the resistance of the first atomic plane in contact with the substrate – *surface* – has to be distinguished from multilayers on top of it. Considering that the screening length inside graphite in $c$-direction is 0.38-0.5 nm (30) the data in Figure 3a indicating a deviation from $R_{ST}\sim1/H$ scaling for $H$ below $n=2$ is reasonable.

We now turn to the analysis of the noise data for graphene multilayers as $H$ changes from that of a graphite film to an ultimate surface. Our sample can be viewed as a parallel resistor network consisting of the surface – SLG in contact with the substrate – and graphene multilayer above it (see insets to Figure 3). The first atomic plane on the substrate with the sheet resistance $R_S$ is connected in parallel with the other $n$-1 layers on top of it. The ($n$-1) layers have the sheet resistance of $R_B$. The $R_B$ value can be extracted from the measured total resistance $R_T$ as $R_B=(R_{ST}\times R_S)/(R_{ST}-R_S)$. Figure 3b shows $R_B$ as a function of ($n$-1) fitted with the power law $R_B=\rho_B/(n-1)^\beta$, where $\rho_B$ is 18.2 kΩ-per atomic plane and $\beta=1.7$. The area-normalized surface noise $S_{RS}/R_S^2$ is generated in the first atomic plane – surface – while the area-normalized volume noise originates in the other ($n$-1) layers. The volume noise scales with the thickness as $1/(n-1)$ (i.e. $1/H$). Both the surface noise from the first graphene atomic plane and the volume noise from multilayers above contribute to the noise measured from the total sheet resistance of $R_{ST}=R_B\times R_S/(R_B+R_S)$.

[Figure 3]

The 1/f noise spectral density $S_I/I^2 = S_R/R_{ST}^2$ includes components $R_B^2/(R_S+R_B)^2 \times S_{R_S}/R_S^2 \times A$ from the surface and $R_S^2/(R_S+R_B)^2 \times S_{R_B}/R_B^2 \times A$ from the volume. The volume noise scales with ($n$-1) and can be written as $S_{R_B}/R_B^2 = \xi/(n-1)$, where the noise amplitude of each layer, $\xi$, is the fitting parameter to the experimental data (the derivation details are in the *Supplementary Information*). The value of the surface resistance $R_S$ and $S_{R_S}/R_S^2 \times A$ are the average values for SLG directly measured as 4.43 k$\Omega$ and 2.3×10$^{-8}$ Hz$^{-1}$×μm$^2$, respectively. The bulk resistance is described as $R_B=\rho_B/(n-1)^\beta$ with the experimentally determined parameters given above. When the thickness reduces down to one atomic layer, the total noise consists of the surface noise only.

Figure 4 shows the experimental data for $S_I(f)/I^2 \times A$ vs. $n$ together with the functional dependence predicted by our model. The best fit is obtained for $\xi$=(1/6)×$S_{R_S}/R_S^2 \times A$ =(1/6)×(2.3×10$^{-8}$) Hz$^{-1}$×μm$^2$, which means the surface noise from the first graphene layer – the surface – is six times larger than the noise from each of the other graphene layers, which constitute the volume, i.e. "bulk", of our sample. When plotted separately, the two noise contributions – originating on the surface and in the volume – intersect at $n$~7. This indicates that the surface noise is dominant in graphene multilayers for $n\leq7$. In thicker samples, the 1/f noise is essentially the volume phenomenon. The situation is different in the high-bias regime shown in the inset. The 1/$n^2$ scaling for graphene multilayers ($n$>2) indicates that the noise is dominated by the surface contributions, which is consistent with the data in the inset to Figure 2c. There is no transition to 1/$n$ scaling at $n$~7. In multilayer graphene samples the resistance of the surface layer decreases with increasing |$V_G$| as more charge carriers are induced electrostatically. This leads to the increased contribution of the bottom surface to the overall current conduction and, therefore, to 1/$f$ noise. As a result, the noise in the multilayers increases with increasing |$V_G$| (Figure 2d) revealing a transition from the bulk to surface noise.

[Figure 4]

**Prospective:** Apart from the fundamental science importance of testing directly the origin of 1/$f$-noise, the obtained results are very important for electronics applications. The progress in information technologies and communications crucially depends on the continuing downscaling of the conventional devices, such as main stream silicon complementary metal-oxide-semiconductor (CMOS) technology. As the feature size of CMOS devices is scaled down to achieve higher speed and packing density, the 1/$f$ noise level strongly increases (*31-34*) and becomes the crucial factor limiting the ultimate device performance and downsizing (*6, 33, 34*). The increase in 1/$f$ noise is detrimental not only for transistors and interconnects in digital circuits but also for the high-frequency circuits such as mixers and oscillators, where 1/f noise upconverts into the phase noise (*6*) and deteriorates the signal-to-noise ratio in the operational amplifiers and analogue-digital – digital-analogue converters. With typical dielectric thicknesses in modern transistors of the order of a nanometer, it is important to understand when 1/$f$ noise is becoming a pure surface noise. Although the exact thickness may vary from material to material, the length scale at which the transistor channel or barrier layer becomes the *surface* from the noise prospective is valuable for design of the next generation electronics. The obtained results are also crucial for the graphene applications. Graphene does not have a band-gap, which seriously impedes its prospects for digital electronics. However, graphene revealed potential for the high-frequency analog communications, interconnects (*35*) and sensors (*36*). These applications require low level of 1/*f* noise, which is up-converted via unavoidable device and circuit non-linearity. Therefore, the knowledge of the scaling law for the 1/*f noise* in graphene multilayers is very important for most of realistic graphene applications in electronics.

**Acknowledgements:** The work at UCR was supported, in part, by the Semiconductor Research Corporation (SRC) and Defense Advanced Research Project Agency (DARPA) through FCRP Center on Functional Engineered Nano Architectonics (FENA) and by the National Science Foundation (NSF) projects US EECS-1128304, EECS-1124733 and EECS-1102074. The work at RPI was supported by the US NSF under the auspices of I/UCRC "CONNECTION ONE" at RPI and by the NSF EAGER program. SLR acknowledges partial support from the Russian Fund for Basic Research (RFBR) grant 11-02-00013.

**Author Contributions:** A.A.B. coordinated the project, supervised graphene device fabrication, contributed to data analysis and wrote the manuscript; G.L. fabricated graphene devices, carried out Raman, AFM and electrical measurements, and performed data analysis; S.R. performed noise measurements and data analysis; M.S.S. performed data analysis.

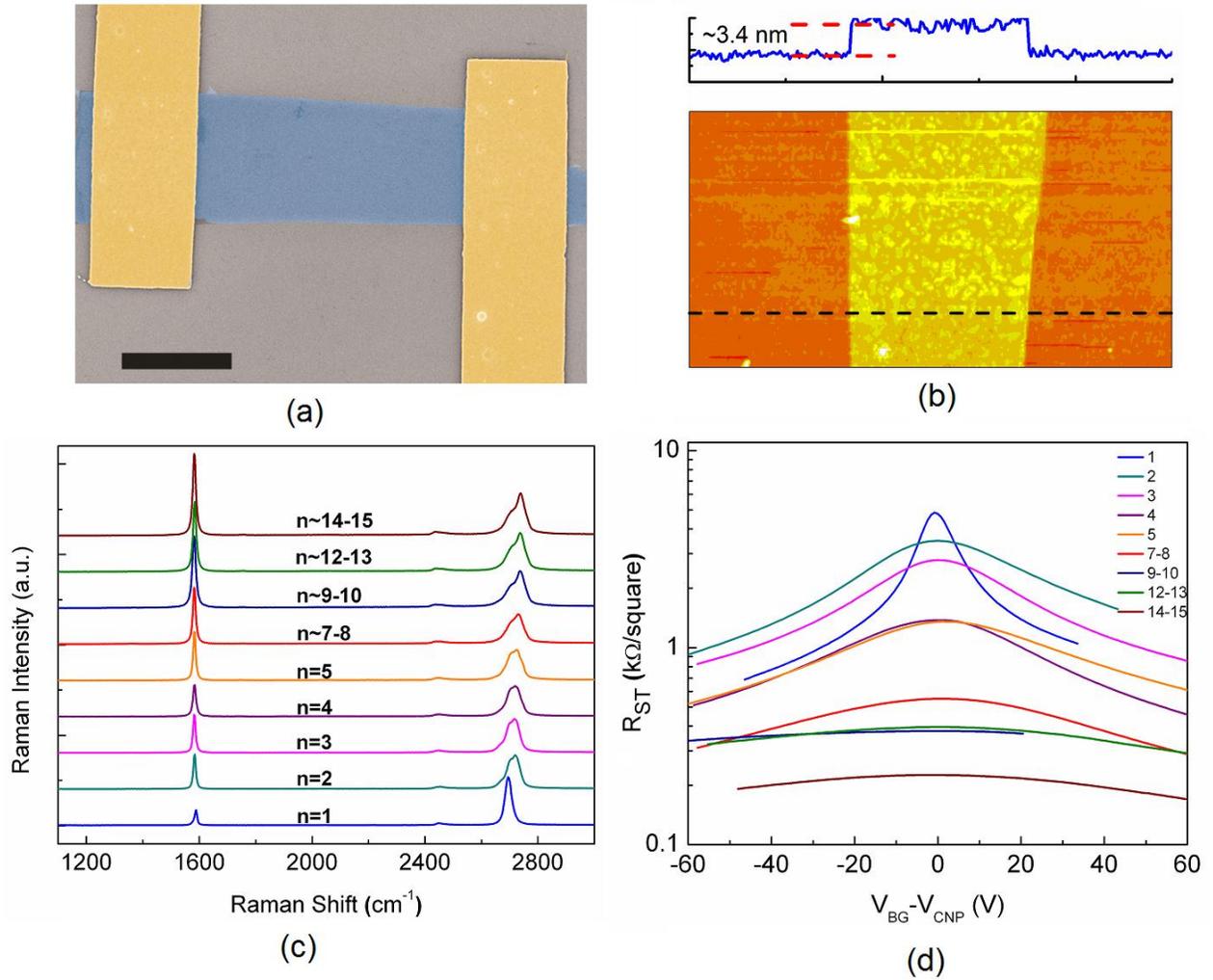

**Figure 1:** (a) Pseudo - color SEM image of a device used for noise measurements showing graphene multilayer (blue) and metal electrodes (yellow). The scale bar is 3 μm. (b) AFM image of a graphene multilayer with the scan direction marked by the dash line. The scanning profile indicates the apparent AFM thickness of 3.4 nm, which corresponds to $n\approx 7$ layers with the carbon-bond thickness $h$=0.35 nm. (c) Raman spectra of the graphene multilayers with different number of atomic planes $n$. The 2D band undergoes noticeable evolution up to $n$=7-9 layers while for the thicker samples the *2D* band becomes indistinguishable from that of bulk graphite. None of the samples reveal *D* peak around 1350 cm$^{-1}$ attesting the sample quality and absence of the fabrication induced defects. (d) Sheet resistance, $R_{ST}$, of graphene and graphene multilayers as a function of the back-gate bias, $V_G-V_{CNP}$, referenced to the charge neutrality point. The data is shown for the continuously changing thickness, $H=h\times n$, from $n=1$ to $n=15$. The gating becomes weaker as $n$ increases.

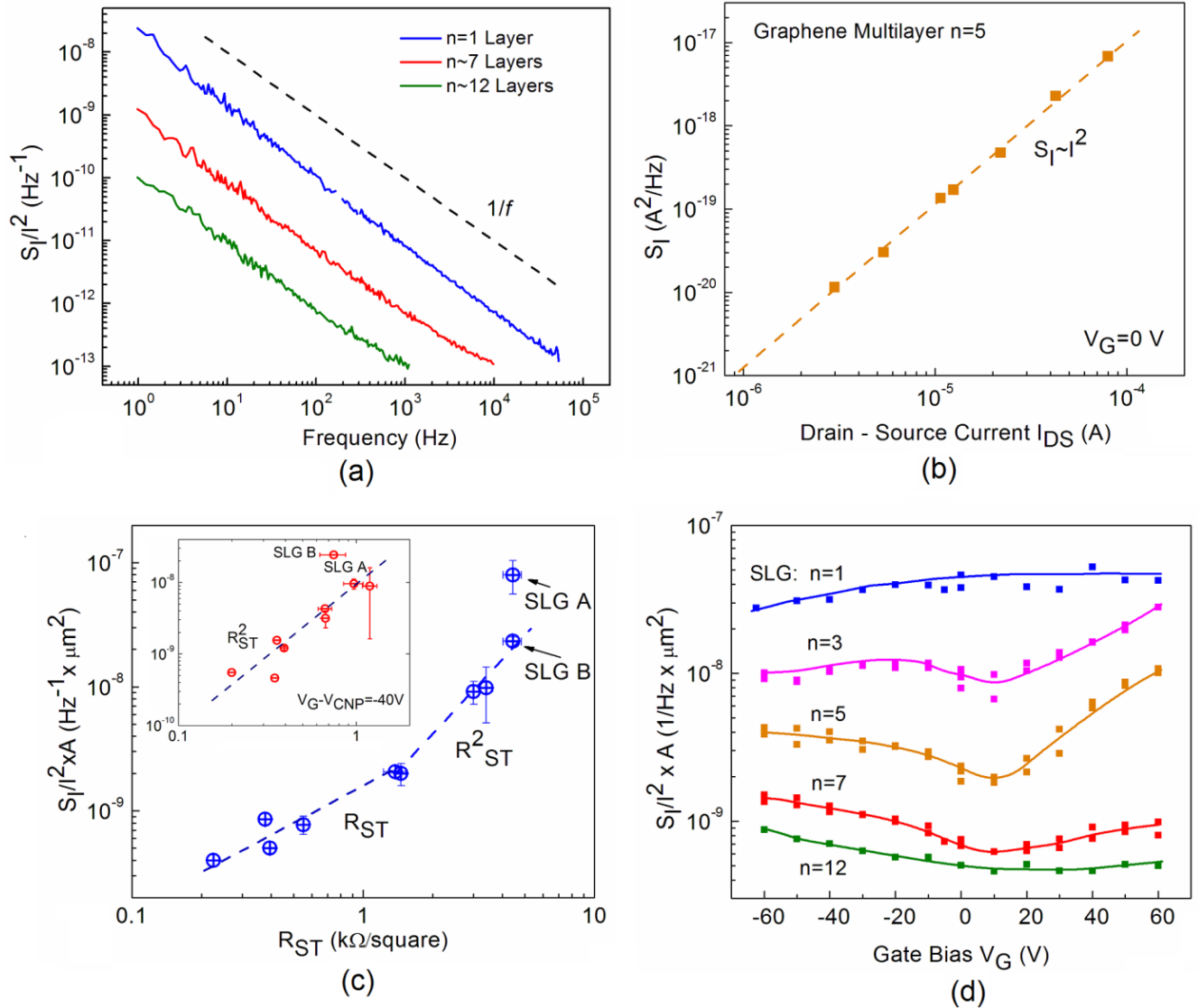

**Figure 2:** (a) Noise spectral density, $S_I/I^2$, of graphene multilayers as a function of frequency shown for three devices with distinctively different thickness: $n=1$, $n\approx7$ and $n\approx12$. In all cases, the noise spectral density is $S_I\sim1/f^\gamma$ with $\gamma\approx1$. (b) Noise density as a function of the drain-source current, $I$, indicating perfect scaling with $I^2$ expected for conventional $1/f$ noise. (c) Noise spectral density normalized by the channel area, $S_I/I^2\times A$, as a function of the sheet resistance of graphene and graphene multilayers. $R_{ST}$ values were taken near the charge neutrality point. Note the noise density scaling with $R_{ST}$ in the low-resistance limit and with $R_{ST}^2$ in the high-resistance limit. Two data points indicated for SLG corresponds to the sample (A) with the thickness $n=1$ over the entire channel and under the metal contact and to the sample (B) with the thickness $n=1$ over the entire channel but increasing to $n=2-3$ under the metal contacts to minimize the contact

noise contribution. Inset shows the noise spectral density as a function of $R_{ST}$ in the high-bias regime ($V_{BG}-V_{CNP}=-40$ V) with the same units. (d) Noise spectral density $S_I/I^2 \times A$ as a function of the gate bias for several multilayers.

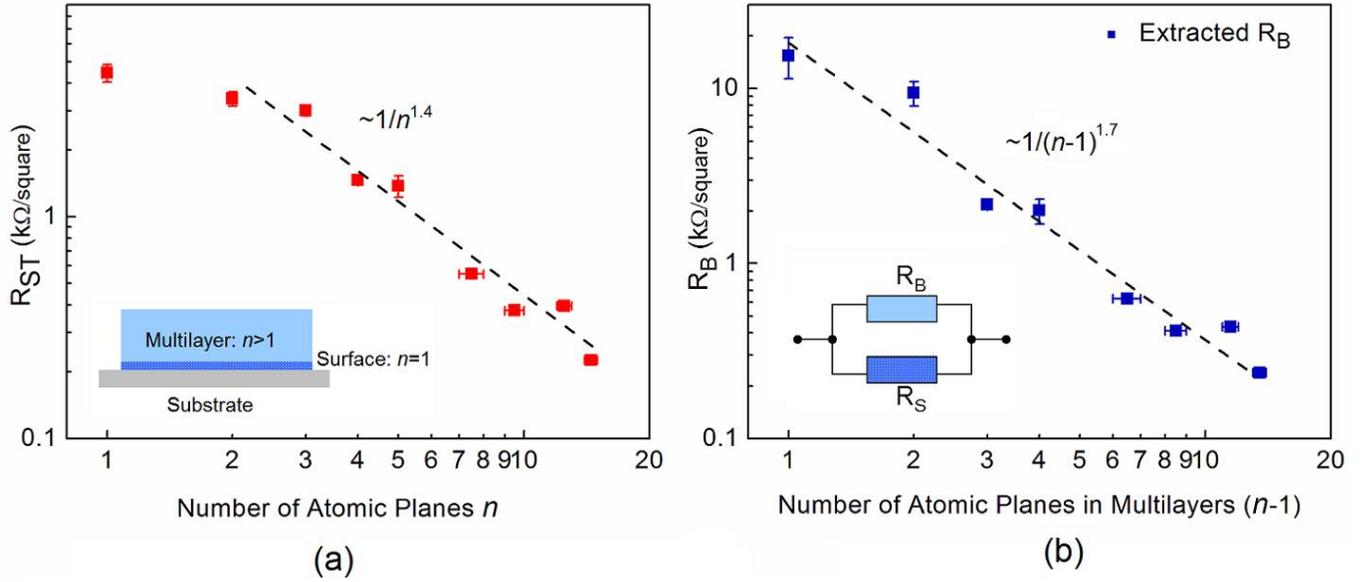

(a)

(b)

**Figure 3:** (a) The measured sheet resistance of graphene and graphene multilayers as a function of the number of atomic planes $n$ near CNP. (b) The extracted sheet resistance, $R_B$, of graphene multilayers on top of the first atomic plane – surface – as the function of the number of atomic planes $n$-1. The insets show the parallel resistor network used to model the resistance of graphene multilayers. $R_S$ is the sheet resistance of the first graphene layer in direct contact with the substrate, which represents the resistance of the actual surface. $R_B$ is the sheet resistance of graphene multilayers on top of the first atomic plane. The power-law fitting of $R_B$ is used for the noise data analysis.

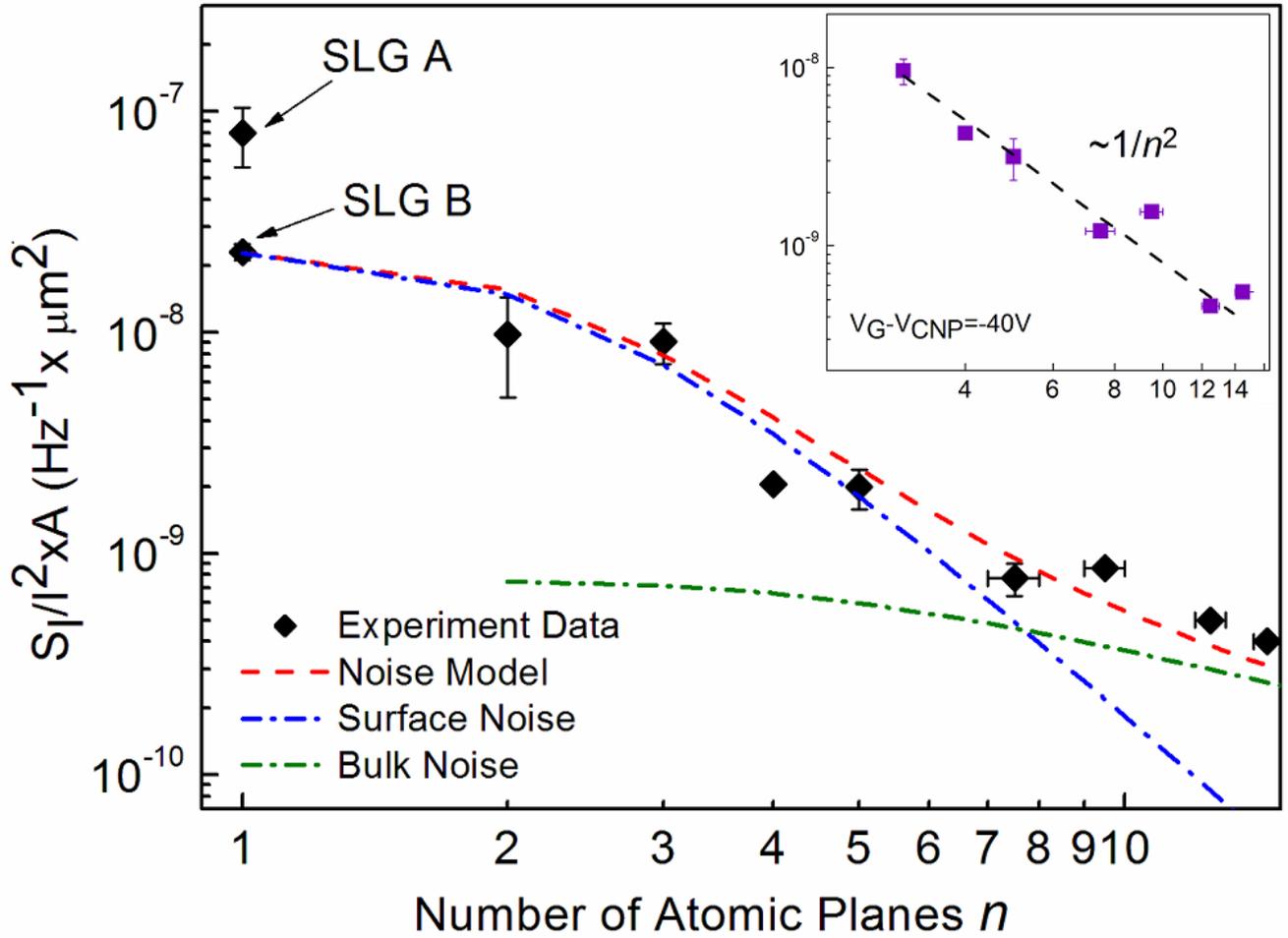

**Figure 4:** The experimental data for the normalized noise spectral density $S_I(f)/I^2 \times A$ near CNP is fitted with the model, which explicitly takes into account the surface noise – originating in the graphene layer in direct contact with the substrate – and the volume (i.e. bulk) noise. For clarity, the surface and volume noise components are also plotted separately. The crossover point at $n \approx 7$ indicates the thickness ($H \approx 7 \times 0.35$ nm $\approx 2.45$ nm) at which the $1/f$ noise becomes essentially the volume phenomenon. In the conducting channels with thickness below this value, the $1/f$ noise is dominated by the surface. The inset shows the noise spectral density in the high-bias regime where surface contributions are more persistent as revealed by $1/n^2$ scaling beyond $n=7$ thickness. The units in the inset are the same as in the main plot.